\documentclass{ws-procs961x669}            
\makeatletter
\renewcommand{\ps@plain}{%
\renewcommand{\@oddhead}{\hfil{\footnotesize%
A contribution to the Julian Schwinger Centennial Conference, %
7--12 February 2018, Singapore}\hfil}%
\renewcommand{\@evenhead}{\@oddhead}%
\renewcommand{\@oddfoot}{\hfil\thepage}%
\renewcommand{\@evenfoot}{\thepage\hfil}%
}
\makeatother
\crop[off]
 
\begin{document}
\title{My Years with Julian Schwinger:\\From Source Theory through 
Sonoluminescence}

\author{K. A. Milton$^*$}

\address{Homer L. Dodge Department of Physics and Astronomy,
University of Oklahoma,\\
Norman, OK 73019, USA\\
$^*$E-mail: kmilton@ou.edu\\
www.nhn.ou.edu/$\sim$milton}



\begin{abstract}
I recall my interactions with Julian Schwinger, first as a graduate student at
Harvard, and then  as a postdoc at UCLA, in the period 1968--81, and 
subsequently.  Some aspects of his legacy to physics are discussed.
\end{abstract}


\bodymatter

\section{The Birth of Source Theory}\label{sec1}
When I came to Harvard in 1967, Julian Schwinger had
already heeded the advice he gave in his
Nobel Lecture to find an alternative to quantum field theory.\cite{nobel}
He was increasingly concerned that conventional field theory, which
he had so very largely developed, was becoming physically remote.
Renormalization was supposed to connect the fundamental fields
with the physical particles observed in the laboratory. His attempts to
include gravity and dual electrodynamics in the framework which so gloriously
accommodated quantum electrodynamics sparked his frustration, as did the 
general
feeling then that quantum field theory could not describe strong interactions.
 Surely, Schwinger had mused, there had to be a more direct way to confront
the phenomena of nature.

So within a year, he did come up with a more phenomenological
approach, which he dubbed {\em source theory.}
Source theory papers started appearing in
1966, \cite{Schwinger:1966zz} first applied to electrodynamics,
\cite{Schwinger:1967rg} and
then to chiral symmetry. \cite{Schwinger:1967tc}
The emphasis was on effective Lagrangians,
and the avoidance of infinities.
In some sense, source theory blended dispersion
relations and field theory.  That is, typical calculations of processes
involving virtual particles involved constructing a situation where real
particles were exchanged between effective sources, and the resulting 
amplitude was ``space-time extrapolated'' to the general situation,
with the amplitude being written as a spectral form.  At least in
simple cases, the process was straightforward and quite effective.

I learned the theory first from the detailed notes of Wu-yang Tsai, taken
the first year I was at Harvard, when I
foolishly did not sit in on Julian's
lectures.  By the end of that year, I had approached the great man and asked
if I could work with him.  To my surprise, he was quite encouraging, so I
worked hard the next three years to justify his faith in me.

\section{What Was It Like to Work with Julian?}
Typically, at Harvard, he had twelve students, at various stages of 
development. He was available only on Wednesdays, after
lunch, first come, first served (advance
booking required).  I recall staying up most of the night
on Tuesdays working feverishly, then getting up at the ungodly hour of 8:30 to
post my name near the top of the list kept by his secretary, who
would arrive at 9:00. If you were near enough to the top,
and Julian returned from lunch early enough, you would receive admittance.
 Once you entered Julian's office in the
afternoon, there was no time pressure, although you didn't want
to appear to be too stupid.  If the phone rang, it was invariably ignored.
Time with the master was unmetered!  All his students had distinct problems,
none of which coincided with what Julian was working on himself, but after
a few minutes of explanation, he would come up with a valuable suggestion.
It might not work, but it would take a week or two to follow the ideas through,
and thirty minutes or so of consultation every week or two was more than 
sufficient to keep progress on track.

Other than these weekly or biweekly meetings, we saw Julian in his classes.
All his students, and some faculty, sat in on
his courses on quantum mechanics and field
theory. (Questions were not encouraged!)  The lectures were like
musical performances, and he held the audience in rapt attention.
Every time he taught a course, the content was completely new, so it was 
advantageous
to attend every reincarnation.  Much original research appeared in 
his lectures, and sometime never anywhere else!  (An example was the first
appearance of the Bethe-Salpeter equation.)

My oral exam, in 1969, on a derivation of the
 Lamb shift via ``sidewise dispersion
relations,''  devolved into a dispute between
Paul Martin and Schwinger on the fundamentals of source theory, but I emerged
unscathed, being able to speak for no more than five minutes.
I was rewarded with a copy of his recent Brandeis lectures. \cite{brandeis}
Even easier was  my defense of my thesis, two years later, largely
 on trace anomalies it turned
out.  It consisted of an excellent lunch at a French restaurant
in Westwood. (I successfully answered Julian's 
single question, on my birthplace.)

\section{The Schwinger's Move to California}
As I was working on stress tensors (which
I still am), Clarice and Julian took a
sabbatical to Japan, a wonderful experience
for them.  There he completed his first volume
of {\it Particles, Sources, and Fields}.\cite{psf1}
When they returned in Fall 1970, Julian
announced his plan to move to UCLA, which was
met with great consternation by his gang of students.
However, he invited his three senior students, me, Lester DeRaad, Jr.,
and Wu-yang Tsai, to accompany him as his assistants (postdocs).
The move was accomplished in February 1971.
(They were greeted by the San Fernando
earthquake!)

\subsection{Why did Julian leave Harvard?}
Of course, he was unhappy with the reception of source theory
at Harvard, by his own former students, Paul Martin and Shelly Glashow,
and others---but this was not the chief reason for relocating.
Julian had become obsessed with exercise, after the death in 1958  of 
Wolfgang Pauli to pancreatic cancer, so he took up
skiing and tennis, and California offered
more opportunities. He could, and did,  even have a
pool.  For years, his assistant at the MIT Radiation Laboratory during
the war, David Saxon, who had become chair of the physics department at 
UCLA (eventually he would become president of the whole University of
California system), had been urging Julian to move to UCLA. He finally
accepted, much to the regret of Boston-bred Clarice.  Saxon and Schwinger
both assumed many students would flock to him, as they had when he joined
Harvard in 1946.  But this was not to be.

\subsection{The Sourcerer's Apprentices}
I arrived at UCLA soon after the Schwingers did, while Lester
and Wu-yang arrived in the summer of 1971. We formed a close research
group at UCLA for several years.
Wu-yang stayed till 1976, when he left for Coral Gables, while Lester left in
1978, although he stayed in Southern California; 
I left the following year for Ohio State University in Columbus, when
my wife was offered a job there in the Dance Department. 

As grad students, we never had any social
contact with the great man.
That changed with our change in status: We
had lunch with him (usually including Bob
Finkelstein and visitors) once a week, and
occasionally were invited to the Schwinger's home in
Bel Air, with its magnificent view of LA.
Julian was always a gracious host.  He was not status conscious,
and had a remarkable ability to listen to others.
An example was the occasion when Julian brought  my
wife's nephew, a sullen teenager,  into the
conversation by an insightful remark about Thelonious Monk.

\section{Interactions with Feynman}
Although Schwinger and Richard Feynman were both living in the Los Angeles
area for more than 20 years, they rarely socialized.
A story I learned from Berge Englert is that once the Feynmans were invited to 
the Schwinger's home, and were having a good time, until
another couple arrived, which spoiled the evening for the
Feynmans.
 When they did meet at conferences they were always very
cordial, and had great mutual respect, which dated back to
the encounters at Shelter Island and the Poconos, during their
parallel development of quantum electrodynamics.

\section{\it Particles, Sources, and Fields}
``If you can't join 'em, beat 'em'' was the motto
to his three-volume exposition of source
theory,\cite{psf1,psf2,psf3} which contains a wealth of information
about fields with arbitrary spin, and included
many detailed calculations in QED.  The second volume even includes
a very complete calculation of the fourth order electron anomalous magnetic
moment, which was first correctly calculated by his student Charlie
Sommerfield.\cite{sommerfield,petermann}
As mentioned, the first volume was written during his sabbatical  in Japan,
while the second and third (unfinished)
volumes were completed at UCLA.  He abandoned this book project just as
he was about to embark on strong interactions, being diverted by his
attempt to understand deep inelastic scattering.
Although we three apprentices diligently
proofread the later volumes, Julian forgot to acknowledge our help in
the published books!

\section{Collaborations with his ``Assistants''}
With the dramatic discovery in 1974 of what is
now called the $J/\psi$ particle, Julian was quick
to come up with an explanation, involving a
``hidden sector,'' maybe ``dyons.'' \cite{Schwinger:1974pr,Schwinger:1975km} 
(The latter was based on his Science article, ``Magnetic
Model of Matter,'' written in 1969.\cite{Schwinger:1969ib})
We joined in with papers on the decay of
$\psi(3.7)$ to $\psi(3.1)$.\cite{Schwinger:1975ps}
His work on ``renormalization group without
renormalization or a group''\cite{Schwinger:1975pnas,Schwinger:1975th}
led to parallel
papers with me,\cite{Milton:1974qx} eventually finding
applications in QCD.\cite{Milton:1996fc}

\subsection{Magnetic charge}
Julian's last papers before the ``source
theory revolution'' were on electric and magnetic charge 
renormalization.\cite{Schwinger:1966zza,Schwinger:1966zzb,Schwinger:1966nj}
He revisited the subject periodically
thereafter, with variations on the $eg = n\hbar c/2$
quantization condition.\cite{Schwinger:1968rq,Schwinger:1975ww}
(He bemoaned the lack of experimental
evidence for magnetic monopoles, ``If only the Price had been 
right.''\cite{price})
This led to the joint ``Dyon-dyon scattering''
paper, with rainbows and glories.\cite{Schwinger:1976fr}
The entirely separate but joint interest of Luis Alvarez \cite{Ross:1973it} 
and Julian Schwinger
in the quest for magnetic charge would eventually lead to the OU search
for magnetic monopoles, which set the best limit on their masses for a decade,
\cite{Kalbfleisch:2003yt} until LHC data became 
available.\cite{Acharya:2016ukt}

\subsection{Deep inelastic scattering}
Julian's last major effort in ``particle
physics''  was his effort to describe the
dramatic scaling phenomena discovered in
deep inelastic scattering \cite{dis1,dis2} without reference to Feynman's 
partons or Gell-Mann's quarks 
of which he had disparaged the naming.\cite{Schwinger:1969ib}
His approach involved double spectral
forms, related closely to the Deser-Gilbert-
Sudarshan representation.
\cite{Schwinger:1975ti,Schwinger:1976ix,Schwinger:1976dw,Schwinger:1977re,
Schwinger:1977rfa}.  We followed on with some explicit related calculations, 
\cite{Tsai:1975tj,
DeRaad:1975jj} but eventually showed that although the double spectral 
representation was
generally valid, it required anomalous spectral regions, which invalidated some of 
the expected behaviors.\cite{Ivanetich:1978pt}

\subsection{QED in background fields}
Julian also revisited the electrodynamics of
particles in strong magnetic fields, harking
back to the famous ``Gauge Invariance and
Vacuum Polarization'' paper,\cite{Schwinger:1951nm} his most cited paper,
and what I regard as the first source theory paper, even though it was written
in 1951.  This work recalled his days at the Rad Lab during World War II,
where he worked out the theory of synchrotron radiation, 
\cite{beta,Schwinger:1949ym}
independently of the Russians.\cite{iwanenko} 
The sequel to Schwinger's 1949 paper  appeared in 1973,
\cite{Schwinger:1973kp}
and collaborative papers with Tsai and Erber
appeared subsequently.\cite{Schwinger:1974rq,Schwinger:1977ba}
Independent papers based on Schwinger's powerful formalism continued
to emerge.\cite{Tsai:1978tr,Milton:1980wd}

\subsection{The triangle anomaly}
Sometimes the interactions were bottom-up.
What is called the axial-vector anomaly was
discovered by Julian in his famous 1951
paper.\cite{Schwinger:1951nm}
We postdocs showed in 1972 that there
were radiative corrections to neutral
pion decay into two photons, in apparent
(but not actual?) disagreement with the
Adler-Bardeen theorem.\cite{Deraad:1973ee,Milton:1973jk}
Julian later confirmed our result, a
correction by a factor of $1+\alpha/\pi$, but instead
of seeking accommodation with received
wisdom,  wrote a joint confrontational paper.
Before the paper was submitted, Julian gave
a talk on the subject at MIT, which was
badly received.
The remains of that unpublished paper exist
in the 3rd volume of {\it Particles, Sources, and Fields}.\cite{psf3}

\subsection{Supersymmetry}
With the emergence of supersymmetry, and
especially its local version, supergravity,
Julian regretted he had not thought of it
first, since he had expounded the
multispinor basis of particles with integer
and half-integer spin in {\it Particles, Sources, and Fields}.\cite{psf1}
A command performance by Stan Deser in
1977 led to his own reconstruction of the
theory,\cite{Schwinger:1978ra} 
but with negligible impact (according to
the current Inspire HEP database, this paper received only 12 citations).
The same fate befell the follow-up paper by Bob Finkelstein, Luis Urrutia, and 
me in which we reconstructed supergravity following Schwinger's lead.
\cite{Milton:1978qs}

\subsection{Casimir effect}
Invariably,  Julian put his current research into
his lectures, which of course we always
attended.
He became intrigued with how the Casimir
effect could be understood without the zero point
energy which seemed not to be
present in the source theory approach, inspired, I believe,
by conversations with Seth Putterman.  This lecture
got quickly written up\cite{casimir}
and then, with Lester and myself we rederived the Lifshitz
theory,\cite{dielectrics}
which included the
infamous ``Schwinger prescription,'' concerning how
to treat the thermal corrections, an amazingly ongoing controversy.
We went on to rederive the surprising result, 
first found by Tim Boyer,\cite{Boyer:1968uf}
that the Casimir self-stress on a perfectly conducting spherical
shell is repulsive.\cite{Milton:1978sf}
I've been strongly bound to the
Casimir effect ever since.\cite{Milton:2001yy}

\section{Books}
We have already mentioned his monumental three-volume {\it Particles,
Sources, and Fields.} But his continual lectures led to several other volumes.
\subsection{\it Classical Electrodynamics}
For the first time since the War, Julian taught
graduate electrodynamics at UCLA, and we
assistants sat in.
 We started turning the lecture notes into a book,
and ended up with a contract with W.H. Freeman.
 Julian then began to pay attention, decided it
didn't sound enough like himself, and began a
nonconvergent series of revisions.  He worked on it for
the better part of a decade, abandoning the rewriting when he got
to radiation theory.
 After his death, we turned it into the book which
exists today,\cite{ce} with the publisher cycling
from Addison-Wesley to Perseus and now Taylor\&Francis.

\subsection{\it Understanding Space and Time}
Julian and astronomer George Abell
developed a BBC course with the Open
University (1976--79).  The bulk of the course dealt
with cosmology; Julian's part was to explain relativity to a general
audience.   A ``Robie the Robot'' graced the
Schwinger's living room thereafter.  It aired at a ``good
time'' on BBC2, but
rather obscurely on KCET in Los Angeles, because its
release coincided with that of Carl Sagan's
{\it Cosmos}. Unlike Sagan, Julian was not a
natural performer. ``Not God's gift to presenting'' was how he was described by
his producer Ian Rosenbloom. However, a Scientific American Library volume
{\it Einstein's Legacy}\cite{legacy} did
memorialize this effort, which is full of interesting historical
anecdotes and insights.

\subsection{\it Quantum Mechanics}
For the first time Schwinger began to teach quantum mechanics to undergraduates
at UCLA.  (Of course, at Harvard, half his audience for his graduate quantum
mechanics courses consisted of
 bright Harvard undergrads.)  He taught many subjects
in very original ways: the framework was the ``measurement algebra'' that he
had developed in the early 1950s based on an analysis of Stern-Gerlach 
experiments on
polarized atoms.  Some notes based on these ideas were published rather 
obscurely
(for physicists) at the end of that decade
in the Proceedings of the National Academy of Sciences;
\cite{pnas1,pnas2,pnas3,pnas4,pnas5}
most famous were the Les Houches lectures of 1955, which appeared in print
only in 1970 (together with reprints of some of the PNAS papers).\cite{QKD}  
But the definitive record of his lectures only was
published under the editorship of Berge Englert.\cite{QMSAM} 

More impact, of course, followed from
 Schwinger's Quantum Action Principle, which originated at
the same time.  This was immediately applied to his reformulation (his third) 
of quantum electrodynamics.\cite{Schwinger:1951xk,Schwinger:1953tb}  
We cannot trace 
that profound redevelopment here, but note that a rather accessible volume 
describing his action principle
based largely on his lectures over the years has now appeared.\cite{sqap}

\subsection{\it Electromagnetic Radiation}
It will be recalled that much of the impetus for Schwinger's rapid scaling of
the peak of quantum electrodynamics derived from his wartime work on radar 
theory at the Radiation Laboratory. (Ironically, it was thought then that 
``radiation'' was a word that would not suggest classified work was in 
progress).  Of course, toward
the end of the war, he gave brilliant lectures on his researches to the other 
lab workers, and many of these were transcribed by David Saxon, and 
mimeographed.  There
was supposed to be a published volume of these lectures, 
but that never materialized.  

Many years later, after the successful publication of 
{\it Classical Electrodynamics},\cite{ce}
Alex Chao of SLAC and Chris Caron of Springer persuaded me to turn
the archival materials into a book.  I did so to the best of my ability, and 
supplemented it with some bits of the
Schwinger archive at UCLA which had not found their way into the previous 
textbook.  The resulting volume contains also reprints of a number of papers of
Schwinger on 
waveguides, synchrotrons and synchrotron radiation, and diffraction.\cite{er}

\section{Family and Diversions}
\subsection{My marriage to Margarita Ba\~nos}
Alfredo Ba\~nos, Jr., was a colleague of
Julian's at the Rad Lab during the War.
Alfredo moved to UCLA thereafter,
eventually divorcing his first wife and  marrying Alice (a great scandal at
the time), and they had a
child Margarita. She grew up and became a dancer.
Naturally, Clarice thought to introduce
Margarita, who in 1975 had just returned from 6
years with the Royal Ballet School in
London, to one of Julian's apprentices.
Although I was second choice (Lester DeRaad was first, but, unbeknownst to
Clarice, he was already taken), it did work
out, and this year we  celebrated our 40th
anniversary with a most memorable  trip to Bali after the Singapore
Centennial.

\subsection{V. Sattui Winery}
The Schwingers became significant
investors in this winery when it was relaunched
in 1976.
In those days, the winery and the BBC
efforts were a frequent topic of our lunch
conversations, which ranged widely.
On Julian's death, in 1994, the winery introduced a special
Cabernet Sauvignon to commemorate their famous partner.

\subsection{60th Birthday Celebration}
I helped organize, along with Bob
Finkelstein and Margy Kivelson, his Fest in 1978.  
Julian was rather unhappy about the affair, for he thought
of it as a retirement celebration.  But he later apologized
publicly to me and Margy when he received the Monie Ferst
award at Georgia Tech in 1986.
Dick Feynman gave a wonderful talk at the
banquet for Schwinger's Fest.
He recounted his encounters with Julian at
Los Alamos and Pocono.
``Although we'd come from the ends of the
earth with different ideas, we'd climbed the
same mountain from different
sides.''  His remarks were not included in the
60th birthday volume\cite{tcp1} but were in the one
assembled for Julian's 70th,\cite{tcp2} the Fest then
being a somewhat more low-keyed affair. 

\section{The Later Years}
\subsection{Thomas-Fermi and Humpty-Dumpty}
I left UCLA, as noted above, in 1979, when I went to Ohio State as a Visiting
Associate Professor.  But I didn't resign my semi-permanent position at UCLA
until 1981, when I accepted a regular faculty position at another OSU, this
time in Oklahoma.  Berge Englert became my replacement as Julian's assistant in
November of that year.  He immediately joined Schwinger in renewed explorations
of the Thomas-Fermi model of atoms, which Julian had started to analyze in his
undergraduate quantum mechanics course.\cite{tfm1,tfm2} Lester helped at first,
\cite{tfm3} but then Berge became the chief calculator, and an impressive
series of papers followed.\cite{tfm4, sa1, sa2, sa3, sa4, sa5, sa6}
Englert left UCLA for a position in Munich in 1985, 
but their collaboration continued.
Marlan Scully, who was visiting the University of Munich in 1987 involved them 
in a series of papers questioning whether one can reunite beams of atoms which 
have been separated by a Stern-Gerlach measurement, with an unsurprising 
negative answer;\cite{hd1,hd2,hd3} 
Humpty-Dumpty cannot be put back together again.

\subsection{Cold fusion and sonoluminescence}
In 1989 began one of the most remarkable examples of ``pathological science'' 
\cite{langmuir}
with the announcement by Pons and Fleischman, noted chemists, of the discovery 
of cold fusion. \cite{pf}  Of course, fusion occurs at some very low rate at 
ambient temperature due to quantum tunneling, but they claimed to see 
significant energy released.  The history of this sad 
affair is given in the book by Huizenga.\cite{huizenga} It bears on our story 
because Schwinger, nearly
alone among physicists, took the report seriously and believed he could explain
it. The rejection of his paper by Physical Review Letters led him to resign his
fellowship in the American Physical Society and to demand that the ``source 
theory'' index category be deleted, as he would no longer
use it!  (The journal complied, even though the PACS scheme was 
beginning to be widely used by other 
journals, and the source theory category was of course used by others.)

Englert then helped him get the paper published in Zeitschrift f\"ur 
Naturforschung;\cite{cf1} in spite of negative reviews,
the second paper was published in Zeitschrift f\"ur Physik D, \cite{cf2} 
but accompanied
by an editorial note disavowing any responsibility for the validity of the 
conclusions by the 
journal!  Third and fourth papers were rejected and never published, although 
Schwinger wrote
three short notes to the PNAS.\cite{cfpnas1,cfpnas2,cfpnas3}  
Englert reports that eventually
Schwinger began to doubt whether his theory was entirely correct, 
and doubts about the 
experimental evidence rapidly accumulated.  But until his death, he thought 
there was  something right about the phenomena and his explanation of it.

So Julian turned his attention to another seemingly impossible phenomenon, that
of sonoluminescence.
Again he learned about this from his good friend Seth Putterman.  In 
``single-bubble 
sonoluminescence,'' a tiny bubble of air injected into water and submitted to 
suitable ultrasonic
acoustic vibrations undergoes rapid collapse and expansion, which can persist 
for months.
Near the point of minimum radius (a factor of 10 smaller than the 
maximum) a 
flash of visible light is emitted, carrying a total energy of about 10 MeV.  
For a review, see Ref.~\citenum{brenner}.  The dynamics of the bubble seem 
rather well understood, but the
mechanism for the light emission remains poorly understood to this day.

Julian immediately thought: ``dynamical Casimir effect'': the rapidly changing 
boundary conditions
might convert virtual photons into the real ones seen in the observations.  So 
he proceeded to
write a series of papers, the first two being follow-ups on his first Casimir 
effect paper in 1975.\cite{lmp2,lmp3}  The 
balance of the work appears in a series of short notes in PNAS, his
favorite journal where he could avoid scrutiny by cynical referees. 
\cite{slpnas1,slpnas2,slpnas3,slpnas4,slpnas5,slpnas6,slpnas7}
Unfortunately, he had forgotten, or perhaps never realized, that before
I had left UCLA I had considered the Casimir effect for a dielectric sphere; 
\cite{Milton:1979yx} such was the basis
of his estimates, which he never made very precise.  In fact, at our last 
encounter, at the
annual Christmas party at the Ba\~nos' home in Westwood, to which 
Clarice and Julian invariably
came (Julian's job being to hide the three wise men in the Christmas tree), 
Julian 
suggested we work together to put the theory of sonoluminescence on a firm 
footing.  
But that was not to be.  Two months later, he was diagnosed, like Pauli, with 
pancreatic cancer, and he died in July
1994.  I did go back and work out the theory; unfortunately, the conclusion was
 that the energy
balance was too small by a factor of a million to explain the copious 
production of 
photons seen in sonoluminescence.
\cite{Milton:1996wm,Milton:1997ky,Brevik:1998zs}

\section{Schwinger's Legacy}

Julian Schwinger spent nearly as many years at UCLA as at Harvard; the 
former from 
1946-70, the latter 1971-94.  The contrast seems dramatic: he published 150 
papers in his Harvard years,
many of which were groundbreaking
and founded new fields. (Besides the obvious field theory developments, for 
example, think of the Keldysh-Schwinger formalism.\cite{brownian})
In contrast,  the 80 published during the UCLA period seemed more reactive, 
for instance,  suggesting alternatives
to the Weinberg-Salam-Glashow theory \cite{wboson} or an alternative 
approach to the 
renormalization group \cite{Schwinger:1975pnas,Schwinger:1975th}
 or to supersymmetry,\cite{Schwinger:1978ra} 
 or even wrong-headed as in the cold-fusion papers.\cite{cf1}
One could blame much of this on his heroic attempt to reformulate field theory 
free 
from infinities, his source theory.  Had he not been so confrontational in his 
demand
 that students totally divorce themselves from the conventional approach, but 
recognized that he was developing an effective action approach which
offers numerous computational advantages, reception to his ideas would have 
been much
 more favorable and he would have attracted more students.  (He did invent the
concept of effective action, after all.)

One could argue that the fact he had only some 5 students at UCLA as compared 
to (depending on how
one counts) nearly 80 at Harvard led to his increased isolation.  But maybe it 
was not Schwinger who
changed, but the world around him.  In 1965 conventional field theory, which 
Julian had largely developed,
seemed to have reached a dead end, and current algebra and dispersion relations
seemed the way forward
to understand hadronic physics.  Julian tried to find a third way, but just as 
he was getting launched,
the electroweak theory (for which he had laid most of the groundwork 
\cite{tfi}) was 
shown to be consistent
and experimentally verified.  Source theory might be efficient, but it was not 
necessary, and so was
ignored, since only by contact with the master could you become initiated.  It 
is unfortunate that
Julian gave up his reconstruction of field theory just at the point where he 
was turning to strong
interactions, which was what had impelled him to start this development.  He 
could have contributed,
like Feynman, to the elucidation of the non-Abelian theory of quantum 
chromodynamics,  which is still
more of a framework than a precisely calculable theory like QED.  Instead he 
turned 
to his ``dispersive'' approach to deep inelastic scattering, which led to 
limited insight.

His most cited papers from the 
UCLA years, with over 200 citations each according to INSPIRE,  are those
on the Casimir effect.\cite{dielectrics,Milton:1978sf}.  
His work on this subject lives on, even though
we practitioners embrace the notion that the effect reflects the change of
 zero-point energy or field fluctuations which Julian rejected.  This work has 
technological applications,
and undoubtedly has
 something to say about the accelerating universe we live in.

For much more information about the life and work of Julian Schwinger, please 
see the biography~\citenum{mm}, 
updated in Ref.~\citenum{kam}.  This presentation grew out of 
Ref.~\citenum{kam17}.

\section{Acknowledgements}
I thank Berge Englert and all the other organizers of the Julian Schwinger
Centennial for putting together such a wonderful memorial, and we are
all indebted to the
Julian Schwinger Foundation for financial support of the Conference.  
I am grateful to Berge for his encouragement of my writing of this essay.


\begin{thebibliography}{999}
\bibitem{nobel} J. Schwinger, ``Relativistic Quantum Field Theory,'' in {\it Nobel
Lectures---Physics, 1963--1970} (Elsevier, Amsterdam, 1972).  [Reprinted in {\it
Physics Today}, June 1966.]

\bibitem{Schwinger:1966zz}
  J.~Schwinger,
  Particles and sources,
{\it  Phys.\ Rev.\ }  {\bf 152}, 1219 (1966).
  doi:10.1103/PhysRev.152.1219

\bibitem{Schwinger:1967rg}
  J.~Schwinger,
 Sources and electrodynamics,
{\it  Phys.\ Rev.\ }  {\bf 158}, 1391 (1967).
  doi:10.1103/PhysRev.158.1391


\bibitem{Schwinger:1967tc}
  J.~Schwinger,
  Chiral dynamics,
{\it  Phys.\ Lett.\ }  {\bf 24B}, 473 (1967).
  doi:10.1016/0370-2693(67)90277-8
 
\bibitem{brandeis}
J.~Schwinger, {\it Particles and Sources,} (Gordon and Breach, New York, 1968).

\bibitem{psf1} 
  J.~Schwinger,
  {\it Particles, Sources, and Fields. Vol.~1,}
  (Addison-Wesley, Reading, Mass., 1970).

\bibitem{psf2} 
 J.~Schwinger,
  {\it Particles, Sources, and Fields. Vol.~II,}
(Addison-Wesley,  Reading, MA, 1973).




\bibitem{psf3}
J. Schwinger, {\it Particles, Sources, and Fields. Vol.~III.}
(Addison-Wesley, Advanced Book Classics, 1989) ISBN-13: 978-0738200552.


\bibitem{sommerfield} C. M. Sommerfield, Magnetic dipole moment of the 
electron, {\it Phys.\ Rev.} {\bf 107} 328 (1957). 

\bibitem{petermann} A. Petermann, Fourth order magnetic moment of the electron,
 {\it Helv.\ Phys.\ Acta} {\bf 30}, 407 (1957)
 
\bibitem{Schwinger:1974pr}
  J..~Schwinger,
  Interpretation of a narrow resonance in $e^+$-$e^-$ annihilation,
{\it  Phys.\ Rev.\ Lett.}  {\bf 34}, 37 (1975).
  doi:10.1103/PhysRevLett.34.37
 

\bibitem{Schwinger:1975km}
  J.~Schwinger,
  Speculations concerning the $\psi$ particles and dyons,
{\it  Science} {\bf 188}, 1300 (1975).
  doi:10.1126/science.188.4195.1300


\bibitem{Schwinger:1969ib}
  J.~Schwinger,
  A magnetic model of matter,
{\it  Science} {\bf 165}, 757 (1969).
  doi:10.1126/science.165.3895.757

\bibitem{Schwinger:1975ps}
  J.~Schwinger, K.~A.~Milton, W.-y.~Tsai and L.~L.~DeRaad, Jr.,
  Resonance interpretation of the decay of $\psi'(3.7)$ into 
$\psi(3.1)$,
{\it  Phys.\ Rev.\ D} {\bf 12}, 2617 (1975).
  doi:10.1103/PhysRevD.12.2617

\bibitem{Schwinger:1975pnas}
  J.~Schwinger,
 Photon propagation function: Spectral analysis of its asymptotic form,
{\it  Proc.\ Nat.\ Acad.\ Sci.\ USA}  {\bf 71}, 3024 (1974).

\bibitem{Schwinger:1975th}
  J.~Schwinger,
  Photon propagation function: A comparison of asymptotic functions,
{\it  Proc.\ Nat.\ Acad.\ Sci.\ USA}  {\bf 71}, 5047 (1974).
  doi:10.1073/pnas.71.12.5047
 
\bibitem{Milton:1974qx} 
  K.~A.~Milton,
Spectral forms for the photon propagation function and the Gell-Mann-Low 
function,
{\it  Phys.\ Rev.\ D} {\bf 10}, 4247 (1974).
  doi:10.1103/PhysRevD.10.4247

\bibitem{Milton:1996fc}
  K.~A.~Milton and I.~L.~Solovtsov,
 Analytic perturbation theory in QCD and Schwinger's connection between the
beta function and the spectral density,
{\it  Phys.\ Rev.\ D} {\bf 55}, 5295 (1997).
  doi:10.1103/PhysRevD.55.5295
  [hep-ph/9611438].
 

\bibitem{Schwinger:1966zza}
  J.~Schwinger,
  Electric- and magnetic-charge renormalization. I,
{\it  Phys.\ Rev.}  {\bf 151}, 1048 (1966).
  doi:10.1103/PhysRev.151.1048


\bibitem{Schwinger:1966zzb}
  J.~Schwinger,
  Electric- and magnetic-charge renormalization. II,
{\it  Phys.\ Rev.}  {\bf 151}, 1055 (1966).
  doi:10.1103/PhysRev.151.1055

\bibitem{Schwinger:1966nj}
  J.~Schwinger,
  Magnetic charge and quantum field theory,
{\it  Phys.\ Rev.}  {\bf 144}, 1087 (1966).
  doi:10.1103/PhysRev.144.1087


\bibitem{Schwinger:1968rq}
  J.~Schwinger,
  Sources and magnetic charge,
{\it  Phys.\ Rev.}  {\bf 173}, 1536 (1968).
  doi:10.1103/PhysRev.173.1536



\bibitem{Schwinger:1975ww}
  J.~Schwinger,
  Magnetic charge and the charge quantization condition,
{\it  Phys.\ Rev.\ D} {\bf 12}, 3105 (1975).
  doi:10.1103/PhysRevD.12.3105


\bibitem{price}
P. B. Price,  E. K. Shirk, W. Z. Osborne, L. S. Pinsky.
Evidence for the detection of a moving magnetic monopole,
{\it Phys.\ Rev.\ Lett.} {\bf35}  487 (1975).  doi:10.1103/PhysRevLett.35.487

\bibitem{Schwinger:1976fr}            
 J.~Schwinger, K.~A.~Milton, W.-y.~Tsai, L.~L.~DeRaad, Jr. and D.~C.~Clark,
 Nonrelativistic dyon-dyon scattering,
{\it  Ann.\ Phys.\ (N.Y.)} {\bf 101}, 451 (1976). 

\bibitem{Ross:1973it} 
  R.~R.~Ross, P.~H.~Eberhard, L.~W.~Alvarez and R.~D.~Watt,
Search for magnetic monopoles in lunar material using an electromagnetic 
detector,
  {\it Phys.\ Rev.\ D} {\bf 8}, 698 (1973).
  doi:10.1103/PhysRevD.8.698

\bibitem{Kalbfleisch:2003yt}
  G.~R.~Kalbfleisch, W.~Luo, K.~A.~Milton, E.~H.~Smith and M.~G.~Strauss,
Limits on production of magnetic monopoles utilizing samples from the D0 and
CDF detectors at the Tevatron,
{\it  Phys.\ Rev.\ D} {\bf 69}, 052002 (2004)
  doi:10.1103/PhysRevD.69.052002
  [hep-ex/0306045].

\bibitem{Acharya:2016ukt}
  B.~Acharya,  et al.\ [MoEDAL Collaboration],
  Search for magnetic monopoles with the MoEDAL forward trapping detector in 
13 TeV proton-proton collisions at the LHC,
{\it  Phys.\ Rev.\ Lett.}  {\bf 118},  061801 (2017)
  doi:10.1103/PhysRevLett.118.061801
  [arXiv:1611.06817 [hep-ex]].
 
\bibitem{dis1}
E. D. Bloom, et al., 
 High-energy inelastic $e$-$p$ scattering at 6$^\circ$ and 10$^\circ$
{\it Phys.\ Rev.\ Lett.} {\bf23}, 930 (1969).
Bibcode:1969PhRvL..23..930B. doi:10.1103/PhysRevLett.23.930.

\bibitem{dis2}
M. Breidenbach, et al.,
Observed behavior of highly inelastic electron-proton scattering.
{\it Phys.\ Rev.\ Lett.} {\bf23},  935 (1969).
Bibcode:1969PhRvL..23..935B. doi:10.1103/PhysRevLett.23.935.

\bibitem{Schwinger:1975ti}
  J.~Schwinger,
  Source theory viewpoints in deep inelastic scattering,
{\it  Proc.\ Nat.\ Acad.\ Sci.}  {\bf 72}, 1 (1975)
  [{\it Acta Phys.\ Austriaca Suppl.}  {\bf 14}, 471 (1975)].
  doi:10.1007/978-3-7091-8424-0$_-$9, 10.1073/pnas.72.1.1

\bibitem{Schwinger:1976ix}
  J.~Schwinger,
  Deep inelastic scattering of leptons. Part 1.,
{\it  Proc.\ Nat.\ Acad.\ Sci.\ USA}  {\bf 73}, 3351 (1976).
  doi:10.1073/pnas.73.10.3351

\bibitem{Schwinger:1976dw}
  J.~Schwinger,
  Deep inelastic scattering of charged leptons,
{\it  Proc.\ Nat.\ Acad.\ Sci.\ USA}  {\bf 73}, 3816 (1976).
  doi:10.1073/pnas.73.11.3816

\bibitem{Schwinger:1977re}
  J.~Schwinger,
  Deep inelastic sum rules in source theory,
{\it  Nucl.\ Phys.\ B} {\bf 123}, 223 (1977).
  doi:10.1016/0550-3213(77)90460-6



\bibitem{Schwinger:1977rfa}
  J.~Schwinger,
  Deep inelastic neutrino scattering and pion-nucleon cross-sections,
{\it  Phys.\ Lett.}  {\bf 67B}, 89 (1977).
  doi:10.1016/0370-2693(77)90814-0


\bibitem{Tsai:1975tj}
  W.-y.~Tsai, L.~L.~DeRaad, Jr. and K.~A.~Milton,
  Verification of virtual Compton-scattering sum rules in quantum
electrodynamics,
{\it  Phys.\ Rev.\ D} {\bf 11}, 3537 (1975)
  Erratum: [{\it Phys.\ Rev.\ D} {\bf 13}, 1144 (1976)].
  doi:10.1103/PhysRevD.13.1144, 10.1103/PhysRevD.11.3537

\bibitem{DeRaad:1975jj}
  L.~L.~DeRaad, Jr., K.~A.~Milton and W.-y.~Tsai,
  Deep inelastic neutrino scattering: A double spectral form viewpoint,
{\it  Phys.\ Rev.\ D} {\bf 12}, 3747 (1975)
  Erratum: [{\it Phys.\ Rev.\ D} {\bf 13}, 3166 (1976)].
  doi:10.1103/PhysRevD.13.3166, 10.1103/PhysRevD.12.3747

\bibitem{Ivanetich:1978pt}
R.~J.~Ivanetich, W.~y.~Tsai, L.~L.~DeRaad, Jr., K.~A.~Milton and L.~F.~Urrutia,
 Anomalous spectral regions in source theory,
in {\it Themes in Contemporary Physics (Julian Schwinger's Festschrift)},
eds. S. Deser, H. Feshbach, R. J. Finkelstein, K. A. Johnson, and P. C. Martin,
North-Holland, Amsterdam, 1979, p. 233 [{\it Physica} {\bf96A}, 233 (1979)].

\bibitem{Schwinger:1951nm}
  J.~Schwinger,
  On gauge invariance and vacuum polarization,
{\it  Phys.\ Rev.}  {\bf 82}, 664 (1951).
  doi:10.1103/PhysRev.82.664

\bibitem{beta}
J.~Schwinger, On radiation by electrons in a betatron, 1945 unpublished.  
Transcribed by M. A. Furman, in 
{\it A Quantum Legacy: Seminal Papers of Julian Schwinger}, ed.~K. A. Milton
(World Scientific, Singapore, 2000), pp. 307--331.


\bibitem{Schwinger:1949ym} 
  J.~Schwinger,
  On the classical radiation of accelerated electrons,
{\it  Phys.\ Rev.}  {\bf 75}, 1912 (1949).
  doi:10.1103/PhysRev.75.1912

\bibitem{iwanenko}
D. Iwanenko and I. Pomeranchuk,
On the maximal energy attainable in a betatron,
{\it Phys.\ Rev.} {\bf65}, 343 (1944).

\bibitem{Schwinger:1973kp}
  J.~Schwinger,
  Classical radiation of accelerated electrons. II. A quantum viewpoint,
 {\it  Phys.\ Rev.\ D} {\bf 7}, 1696 (1973).
  doi:10.1103/PhysRevD.7.1696


\bibitem{Schwinger:1974rq}
  J.~Schwinger, W.-y.~Tsai and T.~Erber,
Classical and quantum theory of synergic synchrotron-Cherenkov radiation,
{\it  Ann.\ Phys.\ (N.Y.)} {\bf 96}, 303 (1976)
  [{\it Ann.\ Phys.\ (N.Y.)} {\bf 281}, 1019 (2000)].
  doi:10.1016/0003-4916(76)90194-9

\bibitem{Schwinger:1977ba}
  J.~Schwinger and W.-y.~Tsai,
  New approach to quantum corrections in synchrotron radiation,
 {\it Ann.\ Phys.\ (N.Y.)}  {\bf 110}, 63 (1978).
  doi:10.1016/0003-4916(78)90142-2

\bibitem{Tsai:1978tr} 
  W.-y.~Tsai,
 New approach to quantum corrections in synchrotron radiation. 2.,
{\it  Phys.\ Rev.\ D} {\bf 18}, 3863 (1978).
  doi:10.1103/PhysRevD.18.3863

\bibitem{Milton:1980wd} 
  K.~A.~Milton, L.~L.~DeRaad, Jr. and W.-y.~Tsai,
  Electron pair production by virtual synchrotron radiation,
{\it  Phys.\ Rev.\ D} {\bf 23}, 1032 (1981).
  doi:10.1103/PhysRevD.23.1032



\bibitem{Deraad:1973ee} 
  L.~L.~DeRaad, Jr., K.~A.~Milton and W.-y.~Tsai,
  Second order radiative corrections to the triangle anomaly. I,
{\it  Phys.\ Rev.\ D} {\bf 6}, 1766 (1972).
  doi:10.1103/PhysRevD.6.1766

\bibitem{Milton:1973jk} 
  K.~A.~Milton, W.-y.~Tsai and L.~L.~DeRaad, Jr.,
  Second order radiative corrections to the triangle anomaly. II,
{\it  Phys.\ Rev.\ D} {\bf 6}, 3491 (1972).
  doi:10.1103/PhysRevD.6.3491
  
\bibitem{Schwinger:1978ra} 
  J.~Schwinger,
  Multispinor basis of Fermi-Bose transformations,
{\it  Ann.\ Phys.\ (N.Y.)} {\bf 119}, 192 (1979).
  doi:10.1016/0003-4916(79)90255-0

\bibitem{Milton:1978qs} 
  K.~A.~Milton, L.~F.~Urrutia and R.~J.~Finkelstein,
  Constructive approach to supergravity,
{\it  Gen.\ Rel.\ Grav.}  {\bf 12}, 67 (1980).
  doi:10.1007/BF00756169

\bibitem{casimir}
J. Schwinger, Casimir effect in source theory, {\it Lett.\ Math.\ Phys.}
{\bf 1}, 43 (1975)

\bibitem{dielectrics}
J. Schwinger, L. L. DeRaad, Jr., and K. A. Milton,
Casimir effect in dielectrics,
{\it Ann.\ Phys.\ (N.Y.)} {\bf 115}, 1 (1978).

\bibitem{Boyer:1968uf} 
  T.~H.~Boyer,
 Quantum electromagnetic zero point energy of a conducting spherical shell and 
the Casimir model for a charged particle,
{\it  Phys.\ Rev.}  {\bf 174}, 1764 (1968).
  doi:10.1103/PhysRev.174.1764



\bibitem{Milton:1978sf}
  K.~A.~Milton, L.~L.~DeRaad, Jr. and J.~Schwinger,
 Casimir selfstress on a perfectly conducting spherical shell,
{\it  Ann.\ Phys.\ (N.Y.)} {\bf 115}, 388 (1978).
  doi:10.1016/0003-4916(78)90161-6

\bibitem{Milton:2001yy} 
  K.~A.~Milton,
 {\it The Casimir Effect: Physical Manifestations of Zero-Point Energy,}
 (World Scientific, Singapore,  2001) 301 pp
  doi:10.1142/4505



\bibitem{ce}
J. Schwinger, L. L. DeRaad, Jr., K. A. Milton, and W.-y. Tsai,
{\it Classical Electrodynamics} (Perseus/Westview, New York, 1998)
ISBN-13: 978-0738200569.

\bibitem{legacy}
J. Schwinger, {\it Einstein's Legacy} (Scientific American Books, New York, 
1986).

\bibitem{pnas1} J. Schwinger, The algebra of microscopic measurements, 
{\it Proc.\ Natl.\ Acad.\ Sci.\ USA} {\bf 45}, 1542 (1959).

\bibitem{pnas2} J. Schwinger, The geometry of quantum states, 
{\it Proc.\ Natl.\ Acad.\ Sci.\ USA} {\bf 46}, 257 (1960).

\bibitem{pnas3} J. Schwinger, Unitary operator bases, 
{\it Proc.\ Natl.\ Acad.\ Sci.\ USA} {\bf 46}, 570 (1960).

\bibitem{pnas4} J. Schwinger, Unitary transformations and the action principle,
{\it Proc.\ Natl.\ Acad.\ Sci.\ USA} {\bf 46}, 883 (1960).

\bibitem{pnas5} J. Schwinger, The special canonical group, 
{\it Proc.\ Natl.\ Acad.\ Sci.\ USA} {\bf 46}, 1401 (1960).

\bibitem{QKD} J. Schwinger, {\it Quantum Kinematics and Dynamics} 
(Benjamin, New York, 1970).

\bibitem{QMSAM} J. Schwinger, 
{\it Quantum Mechanics: Symbolism of Atomic Measurements}, ed.~B.-G.
Englert (Springer, Berlin, 2001).


\bibitem{Schwinger:1951xk} 
  J.~Schwinger,
  The theory of quantized fields. 1.,
{\it  Phys.\ Rev.}  {\bf 82}, 914 (1951).
  doi:10.1103/PhysRev.82.914

\bibitem{Schwinger:1953tb} 
  J.~Schwinger,
  The theory of quantized fields. 2.,
{\it  Phys.\ Rev.}  {\bf 91}, 713 (1953).
  doi:10.1103/PhysRev.91.713

\bibitem{sqap}
K.~A.~Milton, {\it Schwinger's Quantum Action Principle} 
(Springer Briefs in Physics, Cham, 2015).

\bibitem{er} K. A. Milton and J. Schwinger, 
{\it Electromagnetic Radiation: Variational Methods, Waveguides and 
Accelerators} (Springer, Berlin, 2006).

\bibitem{tcp1} 
S. Deser, H. Feshbach, R. J. Finkelstein, K. A. Johnson, and P. C. Martin,
eds., {\it Themes in Contemporary Physics}, (Julian
Schwinger's 60th birthday Festschrift) (North-Holland, Amsterdam, 1979).
 [{\it Physica} {\bf96A} (1979)].


\bibitem{tcp2} S. Deser and R. Finkelstein, 
{\it Themes in Contemporary Physics II} 
(Julian Schwinger's 70th birthday Festschrift) (World Scientific, Singapore, 
1989).

\bibitem{tfm1}
J.~Schwinger, Thomas-Fermi model: The leading correction, {\it Phys.\ Rev.\ A} 
{\bf 22}, 2353 (1980).



\bibitem{tfm2}
  J.~Schwinger,
  Thomas-Fermi model: The second correction,
{\it  Phys.\ Rev.\ A} {\bf 24}, 2353 (1981).
  doi:10.1103/PhysRevA.24.2353

\bibitem{tfm3}
L. DeRaad, Jr. and J.~Schwinger,
 New Thomas-Fermi theory: A test, {\it Phys.\ Rev.\ A} {\bf 25},2399 (1982).


\bibitem{tfm4} B.-G.~Englert and J.~Schwinger,
Thomas-Fermi revisited: The outer regions of the atom,
{\it Phys.\ Rev.\ A} {\bf 26}, 2332 (1982).

\bibitem{sa1}
  B.~G.~Englert and J.~Schwinger,
  The statistical atom: Handling the strongly bound electrons,
{\it Phys.\ Rev.\ A} {\bf 29}, 2331 (1984).

\bibitem{sa2}
  B.~G.~Englert and J.~Schwinger,
  The statistical atom: Some quantum improvements,
{\it Phys.\ Rev.\ A} {\bf 29}, 2339 (1984).

\bibitem{sa3}
  B.~G.~Englert and J.~Schwinger,
  The new statistical atom: A numerical study,
{\it Phys.\ Rev.\ A} {\bf 29}, 2353 (1984).  

\bibitem{sa4}
  B.~G.~Englert and J.~Schwinger,
  Semiclassical atom,
{\it Phys.\ Rev.\ A} {\bf 32}, 26 (1985).  

\bibitem{sa5}
  B.~G.~Englert and J.~Schwinger,
  Linear degeneracy in the semiclassical atom,
{\it Phys.\ Rev.\ A} {\bf 32}, 36 (1985).  



\bibitem{sa6}
  B.~G.~Englert and J.~Schwinger,
  Atomic binding energy oscillations,
{\it Phys.\ Rev.\ A} {\bf 32}, 47 (1985). 

\bibitem{hd1}
M.~O.~Scully, B.-G.~Englert, and J.~Schwinger,
Spin coherence and Humpty-Dumpty. I. Simplified treatment,
{\it Found.\ Phys.}  {\bf18}, 1045 (1988).

\bibitem{hd2}
M.~O.~Scully, B.-G.~Englert, and J.~Schwinger,
Spin coherence and Humpty-Dumpty. II. General theory,
{\it Z.~Phys.\ D} {\bf10}, 135 (1988).

\bibitem{hd3}
M.~O.~Scully, B.-G.~Englert, and J.~Schwinger,
Spin coherence and Humpty-Dumpty. III. The effects of observation,
{\it Phys.\ Rev.\ A} {\bf40}, 1775 (1989).

\bibitem{langmuir} I. Langmuir, lecture given in 1953, published in 
{\it Physics Today}, October 1989, p.~36.

\bibitem{pf} M. Fleischmann, S. Pons, and M. Hawkins, Electrochemically induced
nuclear fusion of deuterium,
{\it J. Electanal.\ Chem.} {\bf 261}, 301 (1989)

\bibitem{huizenga} J.~H.~Huizenga, 
{\it Cold Fusion: The Scientific Fiasco of the Century}
(Oxford University Press, 1993). 

\bibitem{cf1} J.~Schwinger, Cold fusion: A hypothesis,
{\it Z. Nat.\ Forsh.\ A} {\bf45}, 756 (1990).

\bibitem{cf2} J.~Schwinger, Nuclear energy in an atomic lattice,
{\it Z. Phys.\ D} {\bf 15}, 221 (1990).

\bibitem{cfpnas1} J. Schwinger, Phonon representations, 
{\it Proc.\ Natl.\ Acad.\ Sci.\ USA}
{\bf 87}, 6983 (1990).

\bibitem{cfpnas2} J. Schwinger, Phonon dynamics, 
{\it Proc.\ Natl.\ Acad.\ Sci.\ USA} {\bf 87}, 8370 (1990).

\bibitem{cfpnas3} J. Schwinger, Phonon Green's functions, 
{\it Proc.\ Natl.\ Acad.\ Sci.\ USA}
{\bf 88}, 6537 (1991).



\bibitem{brenner}
M. P. Brenner, S. Hilgenfeldt, and D. Lohse,
Single-bubble sonoluminescence,
{\it Rev.\ Mod.\ Phys.} {\bf74}, 425 (2002).

\bibitem{lmp2} J. Schwinger, Casimir effect in source theory II,
{\it Lett.\ Math.\  Phys.} {\bf 24}, 59 (1992).

\bibitem{lmp3} J. Schwinger, Casimir effect in source theory III,
{\it Lett.\ Math.\  Phys.} {\bf 24}, 227 (1992).


\bibitem{slpnas1} J. Schwinger, Casimir energy for dielectrics,
{\it Proc.\ Natl.\ Acad.\ Sci.\ USA} {\bf 89}, 4091 (1992).

\bibitem{slpnas2} J. Schwinger, Casimir energy for dielectrics: 
Spherical geometry,
{\it Proc.\ Natl.\ Acad.\ Sci.\ USA} {\bf 89}, 11118 (1992).

\bibitem{slpnas3} J. Schwinger, Casimir light: A glimpse,
{\it Proc.\ Natl.\ Acad.\ Sci.\ USA} {\bf 90}, 958 (1993).

\bibitem{slpnas4} J. Schwinger, Casimir light: The source,
{\it Proc.\ Natl.\ Acad.\ Sci.\ USA} {\bf 90}, 2105 (1993).

\bibitem{slpnas5} J. Schwinger, Casimir light: Photon pairs,
{\it Proc.\ Natl.\ Acad.\ Sci.\ USA} {\bf 90}, 4505 (1993).

\bibitem{slpnas6} J. Schwinger, Casimir light: Pieces of the action,
{\it Proc.\ Natl.\ Acad.\ Sci.\ USA} {\bf 90}, 7285 (1993).

\bibitem{slpnas7} J. Schwinger, Casimir light: field pressure,
{\it Proc.\ Natl.\ Acad.\ Sci.\ USA} {\bf 91}, 6473 (1994).

\bibitem{Milton:1979yx} 
  K.~A.~Milton,
Semiclassical electron models: Casimir selfstress in dielectric and conducting 
balls,
{\it  Ann.\ Phys.\ (N.Y.)}  {\bf 127}, 49 (1980).
  doi:10.1016/0003-4916(80)90149-9

\bibitem{Milton:1996wm} 
  K.~A.~Milton and Y.~J.~Ng,
Casimir energy for a spherical cavity in a dielectric: Applications to 
sonoluminescence,
{\it  Phys.\ Rev.\ E} {\bf 55}, 4207 (1997)
  doi:10.1103/PhysRevE.55.4207
  [hep-th/9607186].

\bibitem{Milton:1997ky} 
  K.~A.~Milton and Y.~J.~Ng,
Observability of the bulk Casimir effect: Can the dynamical Casimir effect be 
relevant to sonoluminescence?,
{\it  Phys.\ Rev.\ E} {\bf 57}, 5504 (1998)
  doi:10.1103/PhysRevE.57.5504
  [hep-th/9707122].

\bibitem{Brevik:1998zs} 
  I.~H.~Brevik, V.~N.~Marachevsky and K.~A.~Milton,
Identity of the van der Waals force and the Casimir effect and the irrelevance 
of these phenomena to sonoluminescence,
{\it  Phys.\ Rev.\ Lett.\ } {\bf 82}, 3948 (1999)
  doi:10.1103/PhysRevLett.82.3948
  [hep-th/9810062].


\bibitem{brownian} J. Schwinger, Brownian motion of a quantum oscillator,
        {\it J. Math.\  Phys.} {\bf 2}, 407 (1961).
\bibitem{wboson} J. Schwinger, How massive is the $W$ particle?, 
{\it Phys.\ Rev.\ D\/} {\bf 7}, 908 (1973).

\bibitem{tfi}   J. Schwinger, A theory of the fundamental interactions,
        {\it Ann.\  Phys.\ (N.Y.)} {\bf 2}, 407 (1957).

\bibitem{mm} J.~Mehra and K.~A.~Milton, {\it Climbing the Mountain:
the Scientific Biography of Julian Schwinger},
(Oxford University Press, 2000).

\bibitem{kam} K.~A.~Milton, In appreciation Julian Schwinger:
From nuclear physics and quantum
electrodynamics to source theory and beyond,
{\it Physics in Perspective} {\bf9}, 70-114 (2007).

\bibitem{kam17} K.~A.~Milton, Reminiscences of Julian Schwinger, 
arXiv:1709.00711, to appear in Berge Englert's new edition
of Julian Schwinger's {\it Quantum Mechanics: Symbolism of Atomic 
Measurements}.

\end{thebibliography}
\end{document}